\documentclass[10pt,aps, prb,reprint,superscriptaddress, showpacs]{revtex4-1}

\usepackage[utf8x]{inputenc}
\usepackage{graphicx}
\usepackage{amsmath,amssymb}
\usepackage{color}
\usepackage{natbib}

\usepackage{ulem}

\newcommand{\trace}[2]{\text{Tr}_{#1}\left[#2\right]}
\newcommand{\fig}[1]{Fig.~\ref{#1}}
\newcommand{\SEC}[1]{Sec.~\ref{#1}}
\newcommand{\APP}[1]{App.~\ref{#1}}
\newcommand{\nn}{\nonumber}

\newcommand{\projector}[1]{|{#1}\rangle\langle{#1}|}

\begin{document}
\title{Analysis of the conditional average and conditional variance of dissipated energy in the driven spin-boson model}

\author{Philip Wollfarth}
\affiliation{Institut f\"ur Theorie der Kondensierten Materie, Karlsruher Institut f\"ur Technologie, 76128 Karlsruhe, Germany}

\author{Yasuhiro Utsumi} 
\affiliation{Department of Physics Engineering, Faculty of Engineering, Mie University, Tsu, Mie, 514-8507, Japan}

\author{Alexander Shnirman}
\affiliation{Institut f\"ur Theorie der Kondensierten Materie, Karlsruher Institut f\"ur Technologie, 76128 Karlsruhe, Germany}


\begin{abstract}
We investigate the conditional average and the conditional variance of dissipated energy considering, as a prototypical example, a driven spin-boson system. We follow a measurement protocol
in which the spin is prepared in a certain initial state before undergoing a periodic driving. Subsequently, the spin is projected onto a post-selected final state. We compare the conditional average of dissipated energy to the lower bound which directly follows from the well known fluctuation relations. We further report
that a special selection of the initial (pre-selected) and final (post-selected) spin states leads to an enhanced energy emission with simultaneous noise suppression at driving times of order of the relaxation time. 
\end{abstract}


\date{\today}

\maketitle

\section{Introduction}
\label{sec:introduction}

Recent developments in control and measurement techniques of mesoscopic quantum circuits are 
expected to open an avenue towards the thermodynamics in the quantum regime~\cite{PekolaNatPhys2015}. 
In the last few years, it turned out that mesoscopic quantum circuits offer suitable tools to 
study modern topics of thermodynamics and statistical physics, such as fluctuation 
relations~\cite{BK77,PhysRevE.60.2721,PhysRevLett.78.2690,RevModPhys.83.771,HusseinPRB2014,HusseinPRB2014_2} and the information thermodynamics~\cite{Sagawa2013,ParrondoNatPhys2015,KoskiPRL2015,KoskiPRL2014,KoskiPNAS2014,KoskiNatPhys2013}. 
It is now well recognized that in quantum systems, basic quantities of thermodynamics such as work have to be 
carefully defined, since they are intimately related to the measurement problem~\cite{RevModPhys.83.771,TalknerPRE2007}. 
A prototypical setup to measure work~\cite{PekolaNJP2013} consists of a driven two-level system, i.e. a qubit, coupled to a bosonic heat bath. 
In this setup, the work is related unambiguously to the amount of heat emitted  to the bath, which works as a calorimeter. 
However, since a single photon emission or absorption process effectively performs a projective measurement of the qubit~\cite{HekkingPRL2013}, 
the coherence would be lost as the number of photons increases. 
Therefore, in order to detect a signature of the quantum coherence in this setup, one would need a high-precision calorimeter to resolve a single photon. 
In the last few years, precise thermometry techniques aiming at a single-photon detection have 
advanced dramatically~\cite{GasparinettiPRA2015,PhysRevApplied.6.024005,Govenius2015}. 

In parallel with these developments, theoretical studies of this setup have also been
advanced~\cite{SchmidtPRB2015,KutvonenPRE2015,BorrelliPRE2015,ViisanenNJP2015,BorrelliPRE2015,PekolaJLTP2016,Suomela2016,PhysRevLett.116.240403}. 
Currently, various effects related to the fluctuation relations are being discussed. 
So, the non-Markovian effect induced by a strong qubit-bath coupling~\cite{SchmidtPRB2015} and that induced by a non-equilibrium subsystem~\cite{KutvonenPRE2015} have been analyzed. 
The effects of incomplete measurements caused by discarding a subsystem~\cite{BorrelliPRE2015} and by a `dark' heat bath~\cite{ViisanenNJP2015} are investigated. 
A finite-size heat-bath is also being considered~\cite{PekolaJLTP2016,Suomela2016}, for a realistic model of a calorimeter. In
the regime of strong coupling driving-induced coherences are reflected in the energy flow~\cite{PhysRevLett.116.240403}.

In our previous work~\cite{PhysRevB.90.165411}, we analyzed this setup from a different point of view. That is, we 
found that, with a proper post-selection, the probability distribution of the dissipated energy contains significant corrections indicating quantum coherence. 
We also demonstrated the quantum version of the detailed fluctuation relation~\cite{Jarzynski_2000}, 
which holds for the probability distribution of dissipated energy $\epsilon$ conditioned by the initial and the final qubit states $|i_S \rangle$ and $|f_S \rangle$.
From experimental point of view, it would be less demanding to measure lower-order cumulants rather than the probability distribution itself. 
In the present paper we focus on the first two cumulants, i.e., the average and the noise. 
We analyze the general structure of the conditional average and variance depending on the choice of initial and final states. 
In particular we find that interesting results occur when the system
is driven off resonance. For certain values of the detuning $\Delta$, the conditional average can even become negative.
Another interesting effect is observed in the case of the pre-selected state being exited $|e\rangle$ while the 
post-selection one being the ground state $|g\rangle$. In this case, we show that, for finite detuning, the 
conditional average of energy reaches its maximum while the conditional variance is minimized.

Although our analysis focuses on the intermediate time scales of order of the relaxation time, most
of the observed affects are attributed to the classical part of the characteristic 
function (for precise definition see Section~\ref{sec:discussion} and  Ref.~\onlinecite{PhysRevB.90.165411}). The time independent part of the latter turns out 
to be very sensitive to the pre- and post-selected spin states.  

Concerning the effect of quantum coherences on the conditional average of the dissipated energy, 
our analysis shows that quantum contributions may still be detectable at elevated temperatures. 
However, these contributions turn out to be almost completely overshadowed by the classical contributions. 
At temperatures well below the driving frequency, $T\ll \omega$,
the quantum contributions become more pronounced.

This paper is organized as follows. In \SEC{sec:model} we briefly describe the system under consideration and 
explain the proposed experimental protocol. 
This is followed in \SEC{sec:example} by an analysis of the first two conditional cumulants of the dissipated energy. In \SEC{sec:discussion} we discuss the results
of the previous two sections and provide further analysis regarding the asymptotic behaviors of the conditional average and variance.
Finally, in \SEC{sec:conclusion} we conclude.

\section{Model and protocol}
\label{sec:model}
The model we are using has already been discussed in Refs. \onlinecite{PhysRevB.90.165411, 1367-2630-16-11-115001}. Nevertheless, we will
briefly review the most important parts. The system under consideration is a periodically driven two-level-system (TLS) which is weakly 
coupled to an external heat bath. The full Hamiltonian is given by $ H(t) = H_S(t) + H_I + H_B$, where $H_B$ is the Hamiltonian of the bath. The system
is transversally coupled to the bath via $H_I=\sigma_x \otimes B$, where $B$ denotes the bath part of the interaction. The Hamiltonian 
of the driven system is given by $H_S(t)=-\frac{\omega_0}{2}\sigma_z + \frac{\Omega_R}{2}\left(\cos(\omega t) \sigma_x - \sin(\omega t)\sigma_y\right)$, where 
$\sigma_i$ are the Pauli-Matrices, $\omega_0$ can be regarded as a static magnetic field in $z-$direction, $\omega$ is the driving frequency and $\Omega_R$
denotes the Rabi-frequency. 

\begin{figure}[t]
\begin{center}
 \includegraphics[width=.95\linewidth]{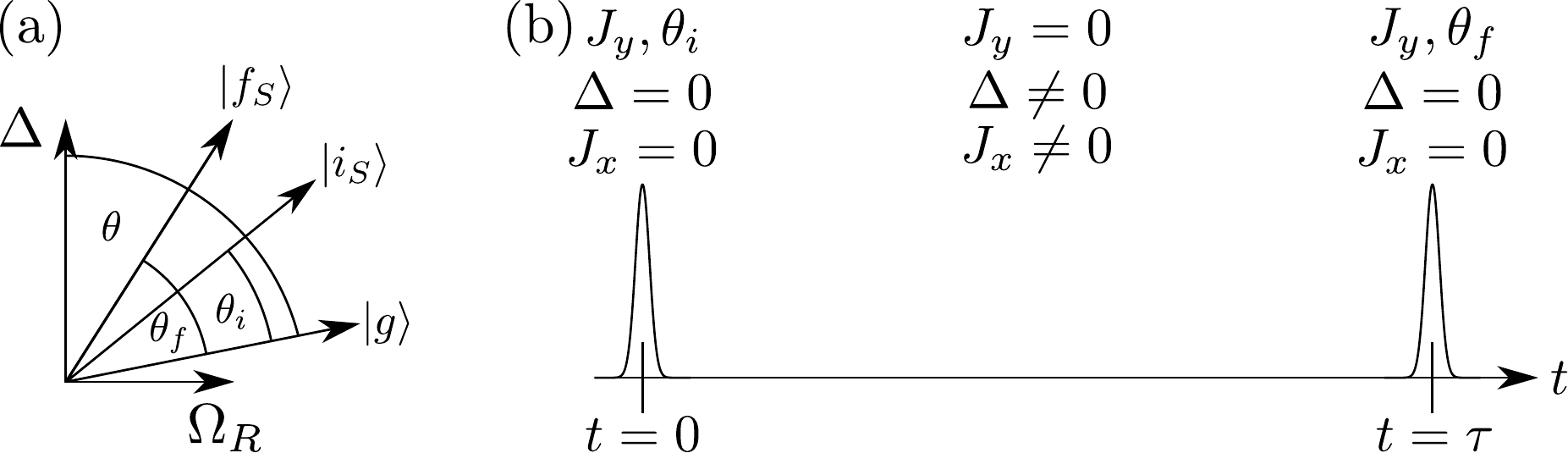}
\end{center}
\caption{Panel (a) shows a schematic describing the necessary angles of the measurement protocol. In
panel (b) the suggested protocol is depicted.}
\label{fig:protocol}
\end{figure}

Transforming the system into the rotating frame yields a time independent Hamiltonian of the driven spin
$\widetilde{H}_S = -\frac{\Delta}{2}\sigma_z +\frac{\Omega_R}{2} \sigma_x$, where $\Delta=\omega_0-\omega$ is the detuning, but shifts 
the periodic time dependency onto the system bath interaction. The calculations are performed in the energy
eigenbasis of the driven spin, which is achieved by a further rotation of the system around the $y$-axis with angle $\theta$. Here $\tan\theta =\Omega_R/\Delta$ (see also \fig{fig:protocol} (a)).

The suggested protocol is schematically shown in \fig{fig:protocol} (b). At time $t=0$ the system is prepared
in a certain initial state $|i_S\rangle$, which is obtained by rotating the ground state of $\widetilde{H}_S$ 
by the angle $\theta_i$ around the $y$-axis as depicted in \fig{fig:protocol} (a). 
This preparation of the initial state may be achieved by a strong resonant $\theta_i$ pulse around the $y$ axis with amplitude $J_y$. 
After the preparation the system is exposed to the possibly off-resonant driving with $J_x=\Omega_R$ and $J_y=0$.
Since changing of the driving frequency $\omega$ may be cumbersome in a realistic experimental
situation, in order to perform off-resonant driving one could adjust the TLS intrinsic energy splitting $\omega_0$. 
At $t=\tau$ the driving is turned off and the system state is post-selected onto the desired final state $|f_S\rangle$
by a second resonant pulse with amplitude $J_y$, inducing a rotating around the $y$-axis by the angle $\theta_f$
and by the subsequent strong measurement. 

The conditional average as well as the conditional variance of dissipated energy are calculated using the method of 
full counting statistics (FCS)\cite{:/content/aip/journal/jmp/37/10/10.1063/1.531672}. More precisely,
in the limit of weak system bath interaction, we adopt the two point measurement approach suggested in Ref. \onlinecite{RevModPhys.81.1665}.
The necessary counting field $\lambda$ is incorporated via $H_\lambda(t) = e^{i \lambda H_B} H(t) e^{-i\lambda H_B}$. The information about
the conditional average of dissipated energy is stored in the characteristic function (CF)
\begin{align}\label{eq:CF}
 \chi_\tau(\lambda,f|i) &= \trace{}{X_f \rho(\lambda,\tau)}
\end{align}
where $\rho(\lambda,t)$ is the counting field dependent density operator of system plus bath and $X_f = \projector{f_S}$ denotes the
projector onto the final state. The CF is connected to the conditional probability distribution via Fourier-transformation
$\chi_\tau(\lambda, f|i) = \int d\epsilon e^{-i\lambda \epsilon} \mathcal{P}_\tau(\epsilon,f|i)$. 

The time evolution of the density operator is derived using a master equation~\cite{breuer2007theory}
\begin{align}\label{eq:MasterEq}
 \frac{d}{dt}\rho_S(\lambda,t) = \mathcal{L}(\lambda)\rho_S(\lambda,t),
\end{align}
where $\mathcal{L}(\lambda)$ denotes the super operator determining the time evolution of the reduced system density matrix $\rho_S(t)$. The
super operator contains the relaxation rates as well as the dephasing rates, which can be found in Ref. \onlinecite{PhysRevB.90.165411}. The 
master equation is of Lindblad-form for $\lambda=0$. For later purposes it is useful to rewrite the generating function
\begin{align}
\label{eq:SolutionME}
 \chi_\tau(\lambda,f|i) = \vec{f}^{\,T} e^{\mathcal{L}(\lambda)\tau}\vec{\rho}_i,
\end{align}
where the time evolution of the density operator is written in the super operator space.
Here, the initial density 
operator $\vec{\rho_i} = \left(\rho_{gg}(0),\rho_{ee}(0),\rho_{eg}(0),\rho_{ge}(0)\right)^T$ and the final state projector
$\vec{f}$ are represented by four-component vectors. 

As shwon previously\cite{PhysRevB.90.165411} the CF splits into 
a classical and a quantum part
$ \chi_\tau(\lambda,f|i ) = \chi_\tau^p(\lambda, f|i) + \delta\chi_\tau(\lambda,f|i)$. 
 The former is determined by the diagonal elements of the 
density matrix whereas the latter by the off-diagonal ones.
This enables a separate analysis of both the classical and quantum contributions to the conditional average.

\section{Investigation of the conditional cumulants}
\label{sec:example}

\subsection{Analysis of the conditional average}
\label{sec:average}
In the following we analyze the conditional average of dissipated energy
\begin{align}
 \langle \epsilon_\tau\rangle_{i\rightarrow f} \equiv \int d\epsilon \,\epsilon\, \tilde{\mathcal{P}}_\tau(\epsilon,f|i),
\end{align}
where $\tilde{\mathcal{P}}_\tau(\epsilon, f|i) = \mathcal{P}_\tau(\epsilon,f|i)/\mathcal{P}_\tau(f|i)$
and $\mathcal{P}_\tau(f|i) =\int d\epsilon \mathcal{P}_\tau(\epsilon,f|i)$. 
We note that the detailed fluctuation relation (FR) directly demands a lower bound on the 
conditional average
\begin{align}\label{eq:second_law}
 \langle \epsilon_\tau\rangle_{i\rightarrow f} \geq \frac{1}{\beta}\ln \left(\frac{\mathcal{P}_\tau(f|i)}{\mathcal{P}_{\tau,B}(i|f)}\right),
\end{align}
which can be understood as the second law of thermodynamics for the pre- and post-selected ensemble.  Here the subscript $B$ indicates the time reversed process (backward protocol). Interestingly, the lower bound for the 
conditional average of dissipated energy can in general be negative, depending on the selection of the initial and final states of the system. 

The parametrization for the initial density operator and the final state projector of the system in the super-operator space is chosen as 
\begin{align}
\vec{\rho_i}&= \left(\cos^2 \frac{\theta_i}{2}, \sin^2 \frac{\theta_i}{2}, \frac{\sin\theta_i}{2},  \frac{\sin\theta_i}{2}\right)^T, \\
\vec{f} &= \left(\cos^2\frac{\theta_f}{2},\sin^2 \frac{\theta_f}{2},\frac{\sin\theta_f}{2},\frac{\sin\theta_f}{2} \right)^T,
\end{align}
such that for $\theta_i=\theta_f=0$ the system will be initially prepared as well as post-selected in the ground state $|g\rangle$ of the system.

\begin{figure}[t]
\begin{center}
\includegraphics[width=\linewidth]{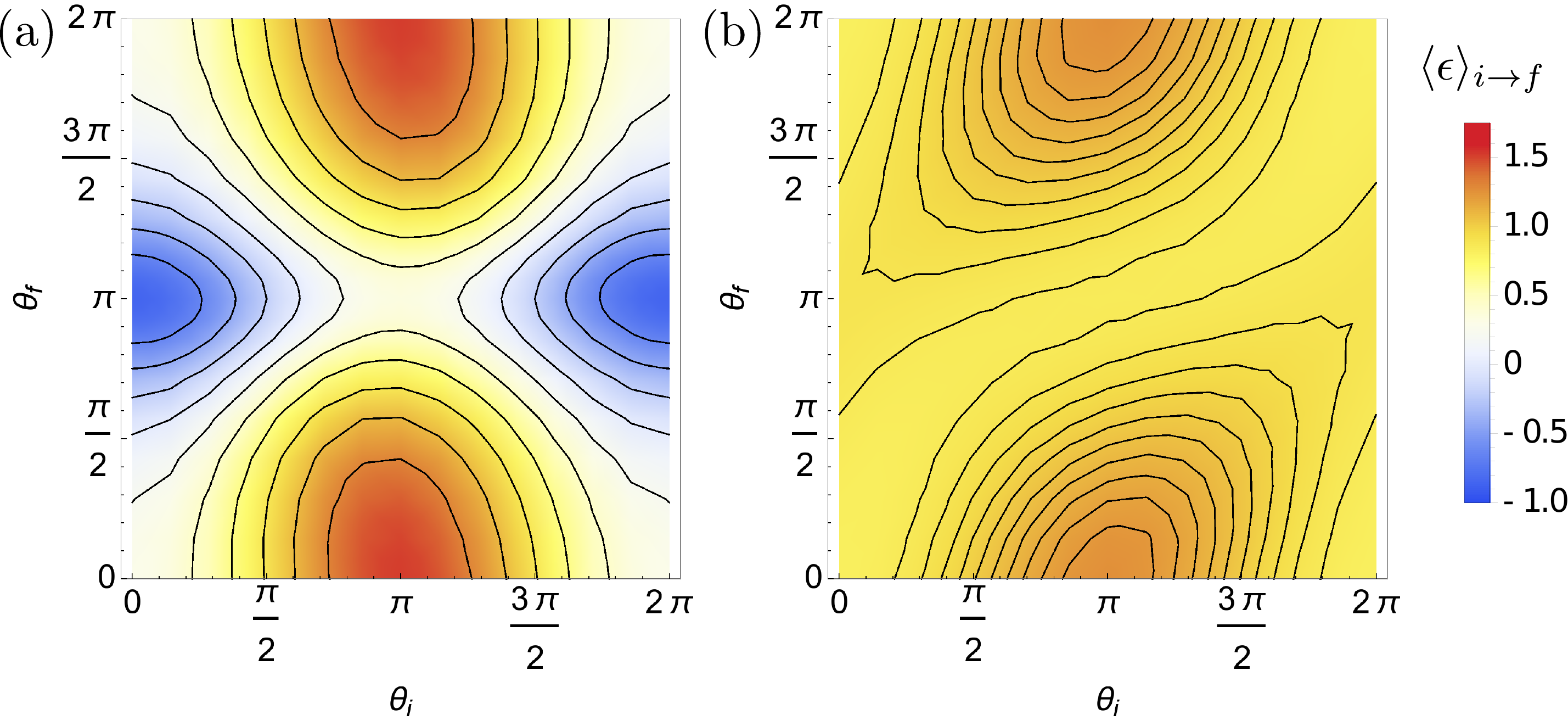}
\end{center}
\caption{(Color online) The conditional average $\langle \epsilon\rangle_{i\rightarrow f}$ depicted at finite temperature $T=\omega$ 
and finite driving time $\omega \tau = 30 \times 2 \pi $.
In panel (a), we have a finite detuning $\Delta = 0.2 \omega$. 
Panel (b) corresponds to the case of resonant driving with $\Delta=0$.
The dimensionless coupling strength between system and bath is set to $\gamma_0=0.01$. The Rabi-frequency is set to $\Omega_R=0.2\omega$.
Lines of equal energy have been included for clarity.}
\label{fig:theta_FI}
\end{figure}

\begin{figure}[b]
\begin{center}
 \includegraphics[width=\linewidth]{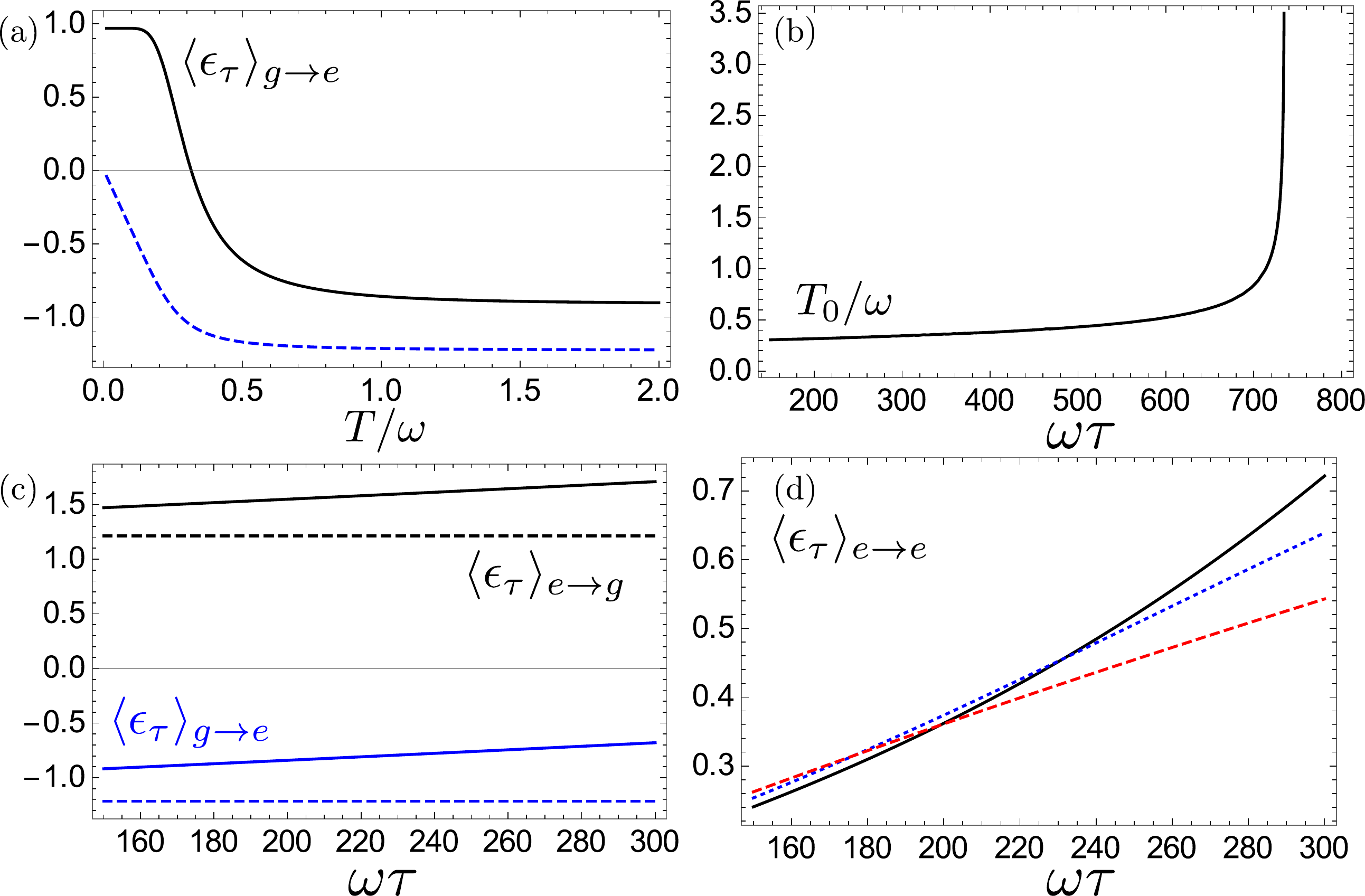}
\end{center}
\caption{(Color online) Conditional average of the energy dissipated to the bath for various choices of the pre- and post-selected states. In all panels $\Delta =\Omega_R= 0.2\omega$.
In panel (a) the conditional average $\langle \epsilon_\tau\rangle_{g\rightarrow e}$ (black) and its lower bound (blue, dashed) is shown 
as a function of the temperature $T$ (driving time $\omega \tau = 30 \times 2 \pi $). Panel (b) shows the transition temperature $T_0/\omega$ for the same
pre and post selection as a function of the driving time $\omega \tau$. In panel (c) and (d) we show the behavior of the conditional average
as function of the driving time. Panel (c) shows the conditional average $\langle \epsilon_\tau\rangle_{e\rightarrow g}$ (black) and
the corresponding lower bound (black, dashed) as well as the the conditional average $\langle\epsilon_\tau\rangle_{g\rightarrow e}$ (blue) and the
lower bound (blue, dashed) respectively. In panel (d) we show the dependence of the conditional average $\langle \epsilon_\tau\rangle_{e\rightarrow e}$ on the driving time for three 
different temperatures: $T=0.01\omega$ (black), $T=0.5\omega$ (blue,dotted) and $T=\omega$ (red, dashed).}
\label{fig:collect_classic}
\end{figure}

In \fig{fig:theta_FI} the conditional average $\langle \epsilon\rangle_{i\rightarrow f}$ is depicted as a function of the 
angles $\theta_i$ and $\theta_f$ for a finite driving time $\omega \tau = 30 \times 2 \pi$ and 
finite temperature $T=\omega$. The driving time is chosen to be of the order (somewhat longer) than the 
characteristic relaxation times of  the spin (the relaxation rates $\Gamma_\text{rel}$ and $\Gamma_\varphi$ 
introduced later are of the order $\Gamma_\text{rel} \sim \Gamma_\varphi \sim 0.015 \omega$). At such times the 
influence of pre- and post-selection is significant. At much longer times the statistics of the dissipated energy 
is dominated by the properties of the stationary state, which establishes in the system irrespective of the initial 
conditions. 
We compare the conditional average in the case of finite detuning, $\Delta=0.2 \omega$, in \fig{fig:theta_FI} (a) and resonant driving, $\Delta=0$, in \fig{fig:theta_FI} (b). 
We observe that in the case of finite detuning the state selection seems to have a significantly larger impact on the amount of energy being dissipated. Additionally, in contrast to the resonantly driven system, 
the case of finite detuning 
exhibits regions, where the conditional average is negative. 
We see that in both cases the conditional average 
reaches its highest value for $\theta_i =\pi$ and $\theta_f = 0$, i.e., when  
the system is initially prepared in its excited state and finally projected onto its ground state (in rotating frame). Interestingly, for this choice of driving time and temperature, the absolute value of $\langle \epsilon\rangle_{e\rightarrow g}$ turns out to be larger at finite detuning as compared to the case of resonant driving.

\begin{figure}[b]
\begin{center}
 \includegraphics[width=\linewidth]{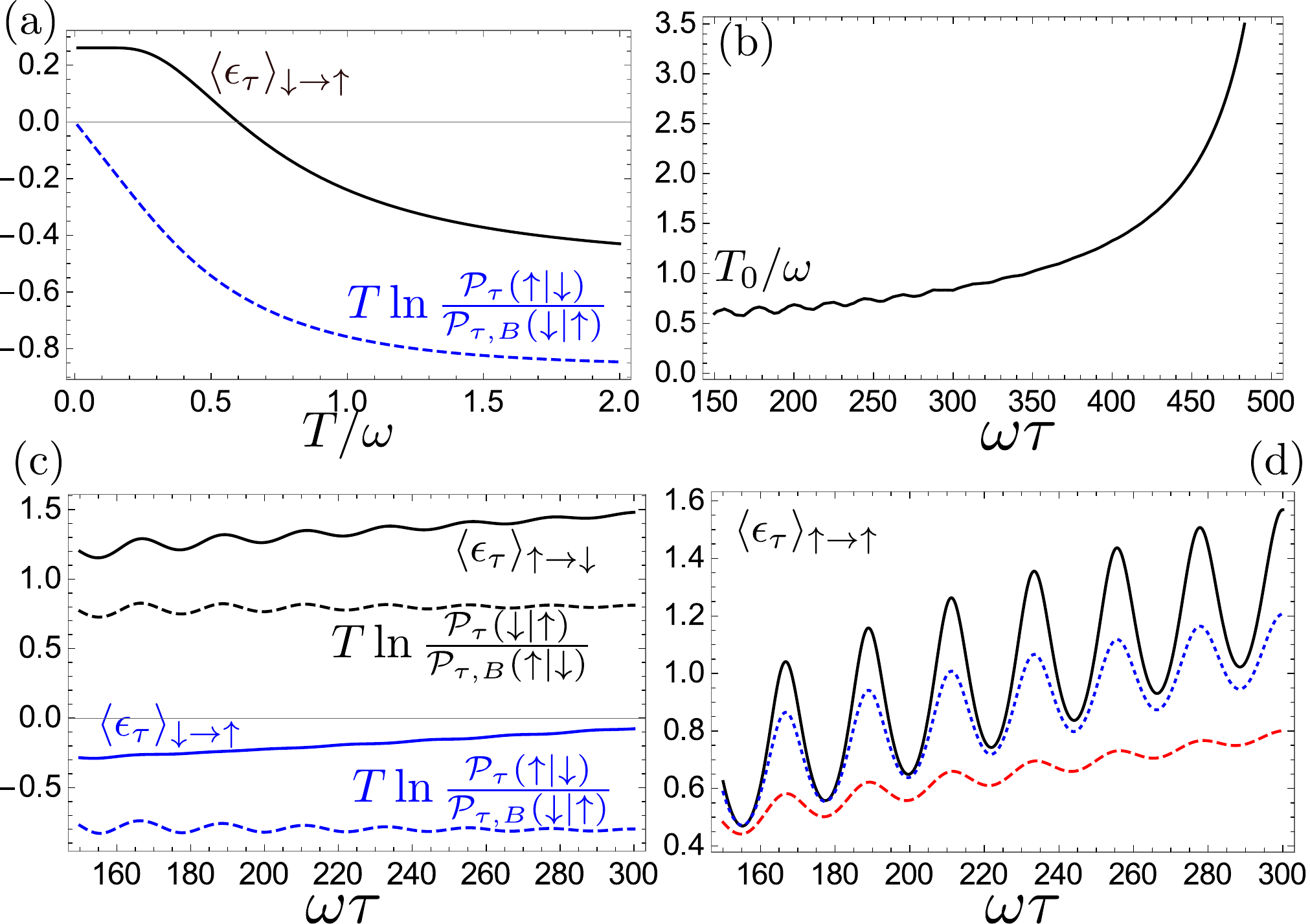}
\end{center}
\caption{(Color online) Conditional average of the energy dissipated to the bath for various choices of the pre- and post-selected states. In all panels $\Delta =\Omega_R= 0.2\omega$. In panel (a) the 
conditional average $\langle \epsilon_\tau\rangle_{\downarrow\rightarrow\uparrow}$ (black) and its lower bound (blue, dashed) (both measured in units of the driving frequency $\omega$) is 
shown as a function of temperature $T$ (driving time $\omega \tau = 30 \times 2 \pi $). Panel (b) shows the transition temperature normalized to
the driving frequency $T_0/\omega$ as a function of the driving time $\omega \tau$. The transition temperature is plotted for the conditional average
$\langle \epsilon_\tau\rangle_{\downarrow\rightarrow\uparrow}$.
Panels (c) and (d) depict the time dependence of the conditional average. In panel (c) we show the conditional averages and the corresponding lower bounds
for different pre- and post-selected states, i.e. $\langle \epsilon_\tau\rangle_{\downarrow\rightarrow\uparrow}$ (blue, bottom) and its lower bound (blue, dashed, bottom)
as well as $\langle \epsilon_\tau\rangle_{\uparrow\rightarrow\downarrow}$ (black, top) and its lower bound (black, dashed, top). Finally panel 
(d) depicts $\langle\epsilon_\tau\rangle_{\uparrow\rightarrow\uparrow}$ as a function of $\omega \tau$  for different 
temperatures $T= 0.01\omega$ (black), $T=0.5\omega$ (blue, dotted) and $T=\omega$ (red, dashed).}
\label{fig:collect}
\end{figure}

In order to achieve a deeper understanding of the results of Fig.~\ref{fig:theta_FI} we consider here certain 
special pairs of pre- and post-selected states. We start (in Fig.~\ref{fig:collect_classic}) from the case of $\theta_i = 0\, {\rm or}\, \pi$ and 
$\theta_f = 0\, {\rm or}\, \pi$, i.e., when both the pre- and the post-selected states are the eigenstates of the 
rotating frame Hamiltonian $\widetilde{H}_S$.  In this case the initial density matrix of Eq.~(\ref{eq:SolutionME})
is purely diagonal and, as discussed in Ref.~\onlinecite{PhysRevB.90.165411}, the off-diagonal elements 
are not generated by the evolution operator. Thus the result is entirely determined by the classical part of the CF. 
In particular no quantum oscillations are expected. 

In \fig{fig:collect_classic} (a)
the conditional average $\langle \epsilon_\tau\rangle_{g \rightarrow e}$, (corresponding $\theta_i=0$ and 
$\theta_f=\pi$) and its respective lower bound, cf. Eq. \eqref{eq:second_law}, are plotted 
as a function of normalized temperature $T/\omega$ for a fixed driving 
time $\omega\tau = 30 \times 2 \pi$. We observe a sign change of the dissipated energy at a transition temperature, which we denote by $T_0$. Above $T_0$ the bath is more likely to 
provide the energy necessary for the transition to the energetically 
unfavorable final state. Below $T_0$ the bath is not capable to excite the system. The system solely receives its energy from the driving source, 
which is partly dissipated to the environment. In \fig{fig:collect_classic} (b) the transition temperature $T_0$ defined above (for $\langle \epsilon_\tau\rangle_{g \rightarrow e}$) is shown as a function of the driving time. For longer times $T_0$ tends to diverge. Indeed, after enough energy has been pumped into the system by the driving source the average dissipated energy to the bath becomes positive, no matter how high the bath temperature itself is.
In \fig{fig:collect_classic} (c) the conditional average and the corresponding lower bound is plotted as a function of the driving time for two different
pairs of pre- and post-selected states. In the case of the energetically unfavorable process 
$|g\rangle\rightarrow|e\rangle$ the 
average dissipated energy is negative and growing (it will become positive at longer times).
We further show the conditional average $\langle \epsilon_\tau\rangle_{e \rightarrow e}$ as a function of $\omega \tau$ for three different temperatures in \fig{fig:collect_classic} (d). As temperature grows the amount of energy dissipated to the bath decreases. Indeed, the bath is more likely to transfer energy back to the system as temperature increases~\cite{1367-2630-17-4-045030}. 

Next we consider the pairs of pre- and post selected states taken from the states $|\uparrow\rangle$ and $|\downarrow\rangle$ (along the $z$-axis in the rotating frame). For $\Delta =\Omega_R= 0.2\omega$, i.e, for $\theta = \pi/4$ the 
pre-selected state $|\uparrow\rangle$ corresponds to $\theta_i = \pi/4$ whereas the pre-selected state $|\downarrow\rangle$ is achieved for $\theta_i = 5\pi/4$ (similarly for the post-selected states and the angle $\theta_f$). 
In Fig.~\ref{fig:collect} we provide results analogous to those of Fig.~\ref{fig:collect_classic}. As the pre- and post-selected states are not the eigenstates of the rotating frame Hamiltonian we observe coherent oscillations that decay due to the relaxation and dephasing processes. We conclude that the qualitative features discussed in relation to \fig{fig:collect_classic} 
remain intact despite the coherent oscillations. Moreover, at high enough temperatures the amplitude of oscillations 
becomes relatively low. 
\begin{figure}[!t]
\begin{center}
 \includegraphics[width=\linewidth]{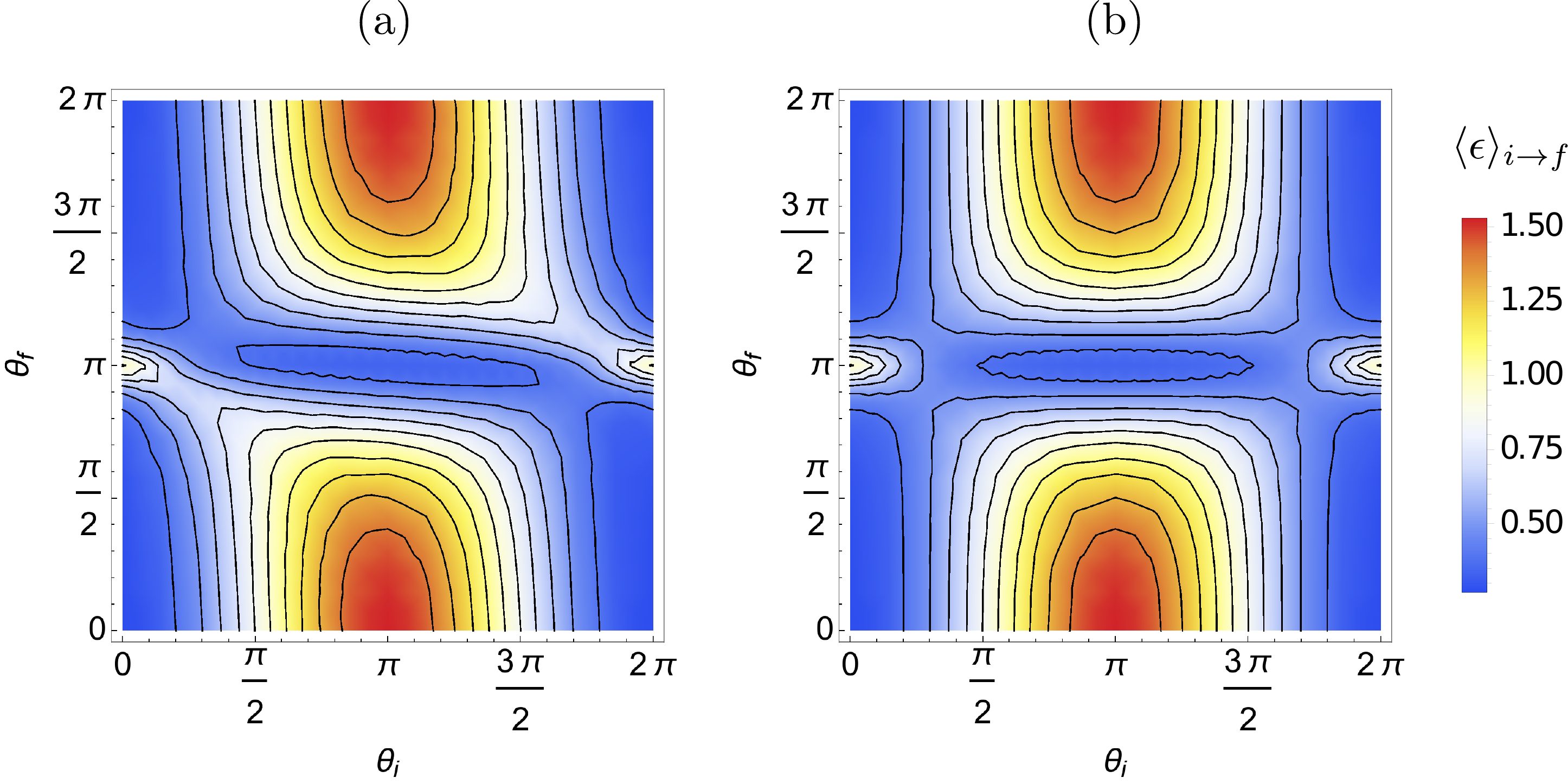}
\end{center}
\caption[Quantum signatures at lower temperatures]
{Contribution of the quantum corrections to the conditional average at low temperatures. In panel (a) we show the conditional average of the dissipated energy where the quantum contributions are fully included. 
In panel (b) the quantum corrections are dropped. The plots are evaluated in the same parameter regime as in Fig. \ref{fig:theta_FI}, except that the temperature is much lower, i.e. $T=0.1\omega$. }
\label{fig:quantumlowtemp}
\end{figure}

Although the most pronounced features concerning the sensitivity to the pre- and post-selection are explained in terms of classical contributions, we analyze
the quantum contribution to the conditional average. It is determined by the quantum part of the CF
$\delta\chi_\tau(\lambda,f|i)$ (see Eq. \eqref{eq:quantum_CF}). 
Consequently, the quantum part of the conditional average is obtained as
\begin{align}\label{eq:quantum_energy}
 \langle \delta \epsilon_\tau\rangle_{i\rightarrow f} &\equiv \frac{\partial_{i\lambda} \left. \delta \chi_\tau(\lambda,f|i)\right|_{\lambda=0}}
 {\mathcal{P}_\tau(f|i)}\nn\\
 &= \frac{ -e^{-\Gamma_\varphi(0)\tau}\omega \tau \cos \Omega \tau }{4 \mathcal{P}_\tau(f|i)}
 \sin^2\theta \sin\theta_i \sin \theta_f  A_B(\omega),
\end{align}
where $A_B(\omega)\equiv \frac{1}{2}(\gamma(\omega)-\gamma(-\omega))$ is the antisymmetrized correlator. The correlation function 
$\gamma(\omega)$
as well as the dephasing rate $\Gamma_\varphi(0)$ are
defined in \APP{sec:appendix_CF}. 
As expected, the quantum features are most noticeable when the pre- and post-selected
states of the system possess maximal coherence, i.e.  $\theta_i=\theta_f= \pi/2$. 
Furthermore, the quantum contributions appear to be largest at driving times of order of the dephasing rate $\tau \sim \Gamma_\varphi(0)$.
As indicated in Fig \ref{fig:collect} (d) the effect of the coherences becomes more pronounced as the temperature decreases. Accordingly, in Fig. \ref{fig:quantumlowtemp}
we show the conditional average as a function of the 
pre- and post-selection angles $\theta_i$ and $\theta_f$ at decreased temperatures $T= 0.1 \omega$.
In Fig. \ref{fig:quantumlowtemp} (a) we show the full conditional average of dissipated energy whereas in Fig. \ref{fig:quantumlowtemp} (b) the quantum corrections are dropped, i.e. 
the conditional average is calculated using Eq. \eqref{eq:ln_chi_p}. Indeed, in the vicinity of the state selection corresponding to maximum coherence, i.e.,  $\theta_i=\theta_f= \pi/2$ and 
$\theta_i = \theta_f = 3\pi/2$, the quantum contributions to the conditional average become visible. 
Note, that we restrict ourselves to the regime of finite detuning $\Delta=0.2\omega$. 

\subsection{Analysis of the conditional variance}
\label{sec:noise}
We further investigate the impact of the pre- and post selection on the variance of dissipated energy
\begin{align}\label{eq:variance}
 \langle \Delta \epsilon^2_\tau \rangle_{i\rightarrow f} &\equiv \langle \epsilon^2_\tau\rangle_{i\rightarrow f} - \langle \epsilon_\tau \rangle^2_{i\rightarrow f}\nn\\
 &= - \partial_\lambda^2 \left.\ln (\chi_\tau(\lambda,f|i))\right|_{\lambda=0}.
\end{align}
The related noise-to-signal ratio, also known as the Fano-factor, has previously been studied in similar
setups, however, regardless the pre- and post selection\cite{PhysRevLett.111.093602,PhysRevE.90.022103}.
\begin{figure}[!t]
\begin{center}
 \includegraphics[width=\linewidth]{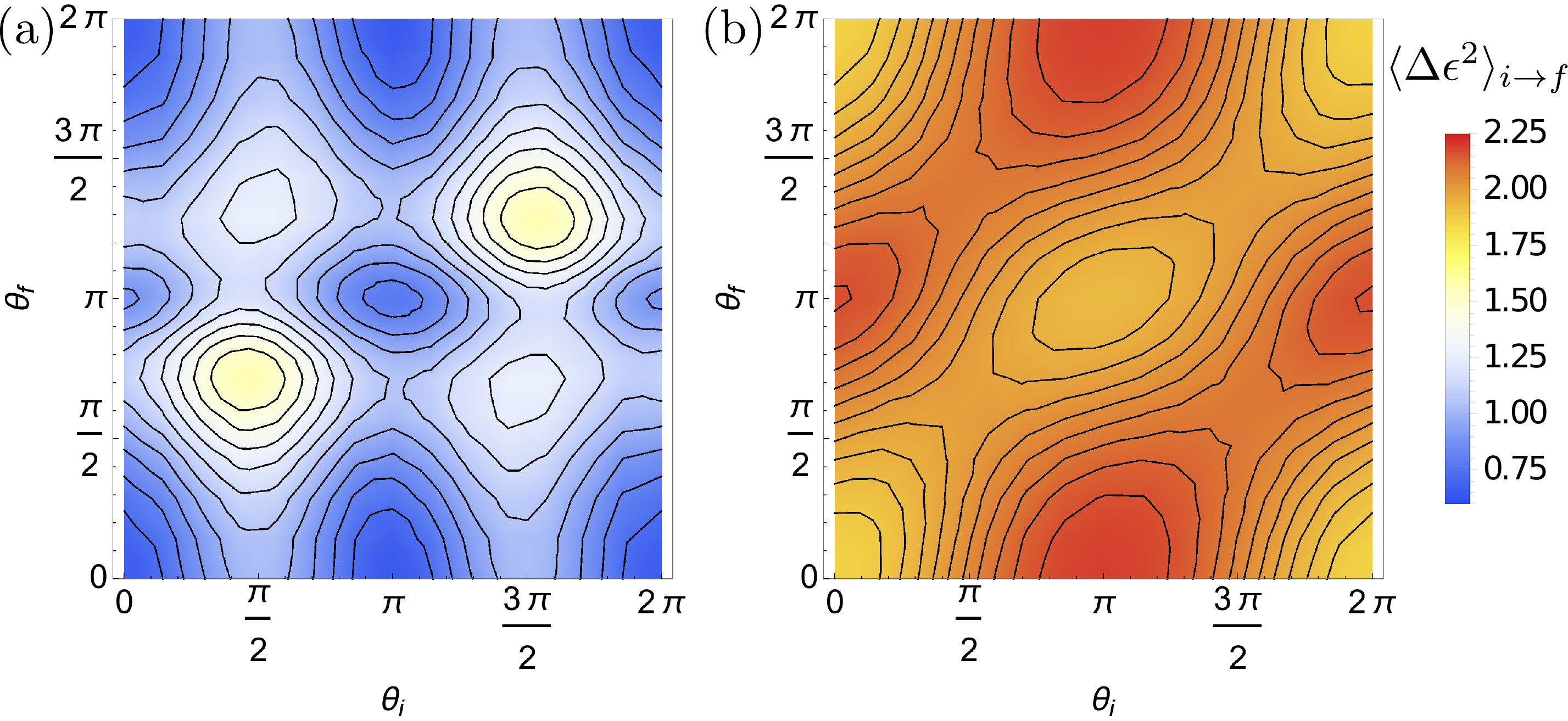}
\end{center}
\caption{(Color online) Conditional variance as a function of the initial and final states characterized by the angles  $\theta_i$ and $\theta_f$. 
The conditional variances are plotted at the time $\omega \tau= 30\times 2\pi$ and temperature $T=\omega$. 
In Panel (a) we consider a finite detuning $\Delta=0.2\omega$. Panel (b) shows the case of resonant driving. 
Lines indicating equal noise amplitude have been included for clarity.}
\label{fig:variance_FI}
\end{figure}

In \fig{fig:variance_FI} we show the results for the conditional variance after the driving time $\omega \tau = 30 \times 2\pi$. 
We investigate the case of finite detuning $\Delta = 0.2\omega$ in \fig{fig:variance_FI} (a), whereas in  \fig{fig:variance_FI} (b) 
we show the resonant case, $\Delta=0$. 
As one could expect the magnitude of the noise is generally larger when the system is driven resonantly.
The variance for the resonantly driven system reaches a maximum for the $|e\rangle \rightarrow |g\rangle$ pre- and post-
selection. Interestingly, in the case of slight detuning \fig{fig:variance_FI} (b), the exact same choice of pre- and post-selected 
system states yields suppressed noise. This will be discussed in the subsequent section.

\section{Discussion}
\label{sec:discussion}
Our main purpose here is to recover the main features of the numerical results shown in \fig{fig:theta_FI}
and \fig{fig:variance_FI} from the classical part of the CF and to explain the qualitative picture behind these features. 
It is reasonable to assume that the classical part of the CF is responsible for most of the observed effects.
One of the reasons is that for driving times of order of the relaxation and dephasing times the 
conditional dissipated energy is strongly influenced by the difference of the energy expectation values in the 
pre- and the post-selected states. The modulation of $\langle \epsilon_\tau\rangle_{i\rightarrow f}$ due to this 
difference should survive even at $\tau\rightarrow \infty$ (it remains, of course, finite and is completely overshadowed 
by the growing with time stationary contributions). Thus the information about the modulation of $\langle \epsilon_\tau\rangle_{i\rightarrow f}$ should be contained in the diagonal elements of the density matrix, i.e., in the classical 
part of the CF. In addition, as we have observed in \fig{fig:collect_classic}  and \fig{fig:collect} the quantum oscillations 
originating in the quantum part of the CF are small at elevated temperatures.

Thus we give here a detailed analysis of the classical CF, which is given by 
\begin{align}\label{eq:ln_chi_p}
 \ln \chi_\tau^p(\lambda, f|i) &= \frac{\tau}{2}(-\Gamma_{gg}(\lambda)-\Gamma_{ee}(\lambda)+ 2\Lambda(\lambda) )\nn\\
 & -\ln 2\Lambda(\lambda) +\ln\frac{A(\lambda,\theta_f,\theta_i)+B(\lambda,\theta_f,\theta_i)}{4} \nn\\
 &+ \ln\left(\!\! 1\!\!-\!\frac{A(\lambda,\theta_f,\theta_i)\!-\!B(\lambda,\theta_f,\theta_i)}{A(\lambda,\theta_f,\theta_i)\!+\!B(\lambda,\theta_f,\theta_i)}e^{-2\Lambda(\lambda)\tau} \!\right)\!.
\end{align}
All the quantities used  here are defined in \APP{sec:appendix_CF}.
The first line of Eq.~(\ref{eq:ln_chi_p}) is related to the eigenvalue of the Liouvillian super-operator $\mathcal{L}(\lambda)$ (cf. Eq.~(\ref{eq:MasterEq})) that vanishes at $\lambda = 0$. This eigenvalue determines 
the long time behavior of the CF (see appendix \ref{sec:appendix_CF}).
The second line of Eq.~(\ref{eq:ln_chi_p}) yields a time independent offset, which carries the information about the 
pre- and post-selected states. The third line of Eq.~(\ref{eq:ln_chi_p}) represents the transient contributions, which 
decay on timescales of order of the relaxation rate $\Gamma_\text{rel}= \Gamma_{eg}+\Gamma_{ge}$. 
Using Eqs. \eqref{eq:chi_p},
\eqref{eq:A} and \eqref{eq:B} it is easy to see that all cumulants vanish at $\tau = 0$.

\begin{figure}[t]
 \begin{center}
 \includegraphics[width=\linewidth]{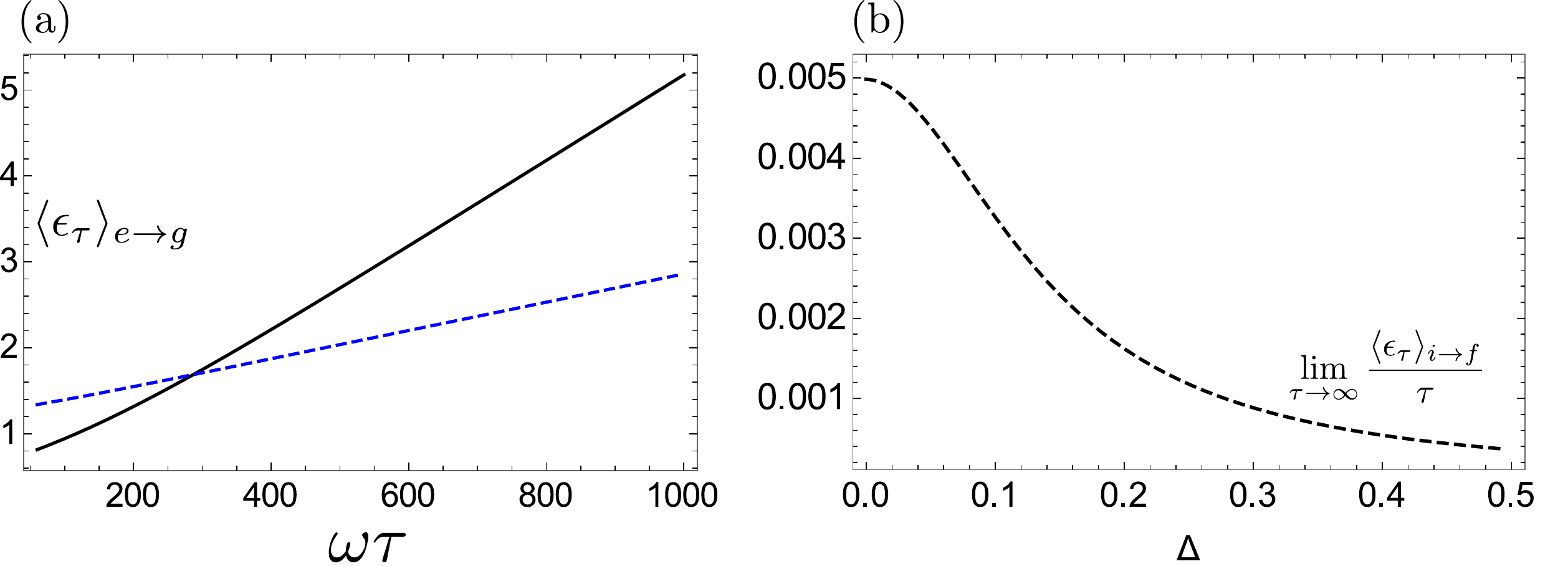}
\end{center}
\caption{(Color online) Conditional average of energy for $|e\rangle\rightarrow |g\rangle$ pre- and post-selection shown in panel (a) for resonant driving
$\Delta=0$ (black, solid) and with finite detuning $\Delta=0.2 \omega$ (blue, dashed). 
Panel (b) depicts the average heat current $\lim_{\tau\rightarrow \infty} \langle \epsilon_\tau\rangle_{i\rightarrow f}/\tau$ as a function of the detuning $\Delta$.}
\label{fig:timeanalysis}
\end{figure}

To establish the relation between the classical part of the CF given by Eq.~(\ref{eq:ln_chi_p}) and 
the complete numerical results of Section~\ref{sec:example} we present
in \fig{fig:timeanalysis} results for the conditional average of the dissipated energy calculated 
with Eq.~(\ref{eq:ln_chi_p}).
In \fig{fig:timeanalysis} (a) the conditional
average  $\langle \epsilon_\tau \rangle_{e\rightarrow g}$ as a function of the driving time both for 
the case of resonant driving as well as for the detuned driving $\Delta =0.2\omega$. 
For short enough driving times more energy is emitted in the detuned case than in the resonant driving case (cf. \fig{fig:theta_FI}). However, the longer the driving lasts, the more energy tends to be dissipated during a resonant drive.

In the long time limit, we find, using Eq.~(\ref{eq:ln_chi_p}), the average heat current of dissipated energy to be given by
\begin{align}
 \lim_{\tau\rightarrow \infty} \frac{\langle\epsilon_\tau\rangle_{i\rightarrow f}}{\tau} &= \frac{\sin^2\theta\,\omega}{4\Gamma_\text{rel}}
 \left(\Gamma_\text{rel} \gamma(\omega) \left(1-e^{-\beta\omega}\right)\vphantom{\frac{1}{2}}\right.\nn\\
 &\left.+ \frac{\sin^2\theta}{2} \gamma(\omega + \Omega) \gamma(\omega-\Omega) \left( 1-e^{-2\beta\omega}\right)\right)\ .
\end{align}
All the quantities used  here are defined in \APP{sec:appendix_CF}. The result is presented in \fig{fig:timeanalysis} (b).
This heat current is completely independent of the pre- and post-selection, but is determined by the detuning of the drive.
As expected, the heat current reaches its maximum for resonant driving and tends to decrease, as detuning increases. 

\begin{figure}[t]
\begin{center}
 \includegraphics[width=\linewidth]{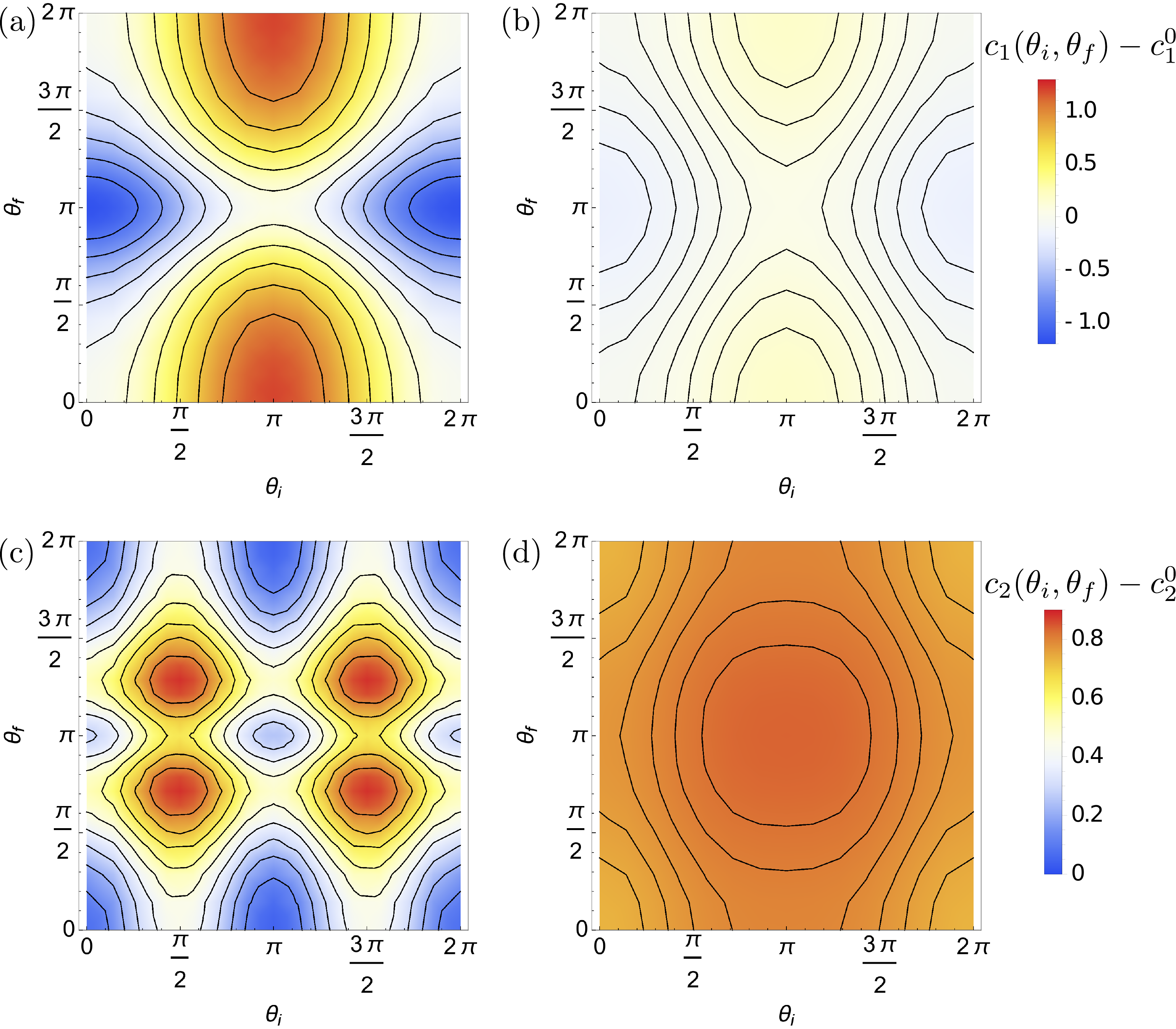}
\end{center}
\caption{(Color online) Top panels: Selection sensitive contribution $ c_1(\theta_i,\theta_f) $ to the conditional average. Panel (a) depicts the situation with $\Delta=0.2\omega$. In 
Panel (b) the situation for resonant driving $\Delta = 0$ is shown.
Bottom panels: Selection sensitive contribution $ c_2(\theta_i,\theta_f) $ to the conditional noise. The panels compare the situation (c) with finite detuning $\Delta =0.2 \omega$ and
(d) resonant driving.}
\label{fig:constant_compare}
\end{figure}

Next, we study the time-independent contributions of the second line of Eq. \eqref{eq:ln_chi_p} to the conditional cumulants of the dissipated energy. These are the leading terms which determine the sensitivity of the cumulants of the dissipated energy to the pre- and post-selected spin states. 
These terms determine the landscape of $\lim_{\tau\rightarrow\infty}\langle \epsilon_\tau\rangle_{i\rightarrow f}$
as a function of $\theta_i$ and $\theta_f$ as all the other contributions dependent of the pre- and post-selection 
vanish at $\tau\rightarrow\infty$.
We drop the first term of the second line of Eq. \eqref{eq:ln_chi_p} as it is independent of $\theta_i$ and $\theta_f$ and define 
\begin{align}
 c_n(\theta_i,\theta_f) \equiv \partial_{i\lambda}^n \left. \ln \frac{A(\lambda,\theta_i,\theta_f) + B(\lambda,\theta_i,\theta_f)}{4}\right|_{\lambda=0}.
\end{align}
The contribution to the  conditional averages (first cumulant) reads
\begin{widetext}
\begin{align}
 c_1(\theta_i,\theta_f) = \frac{(1-\cos \theta_f) (\cos\theta_i+1) \Gamma_{eg}' + (\cos \theta_f +1)(1-\cos\theta_i)\Gamma_{ge}' +\frac{2}{\Gamma_\text{rel}}(1+\cos \theta_f \cos\theta_i)
 (\Gamma_{ge}\Gamma_{eg}'+\Gamma_{ge}'\Gamma_{eg})}{2(\Gamma_\text{rel}+\cos\theta_f(\Gamma_{ge}-\Gamma_{eg} ))}\ .
\end{align}
 \end{widetext}
Here $\Gamma_{ij}'\equiv \partial_{i\lambda} \Gamma_{ij}|_{\lambda=0}$. 
In \fig{fig:constant_compare} we present $c_1(\theta_i,\theta_f)$, 
as a function of the state selection angles for detuned driving (panel (a)) and for the resonant driving (panel (b)). 
We observe a high degree of similarity to \fig{fig:theta_FI}.
In particular, we observe a much stronger dependence on pre- and post-selection in the case of detuned driving 
(cf. \fig{fig:constant_compare} (a) and \fig{fig:theta_FI} (a)) as compared to the regime of resonant driving (cf. \fig{fig:constant_compare} (b) and \fig{fig:theta_FI} (b)).

To explain the higher sensitivity to the pre- and post-selection in the regime of detuned driving 
we analyze the specific choice $|e\rangle\rightarrow |g\rangle$ ($\theta_i = \pi$, $\theta_f=0$) in more detail.
We obtain 
\begin{align}
c_1(\pi,0) &= \frac{\Gamma_{ge}'}{\Gamma_{ge}}\nn\\
 &= \omega+\Omega - \frac{2 \omega \gamma(\omega-\Omega)}{\gamma(\omega-\Omega)+e^{\beta (\omega-\Omega)}\cot^4\frac{\theta}{2} \gamma(\omega+\Omega)}.
\end{align}
That is, the selection of $\theta_i = \pi$ and $\theta_f=0$ identifies the processes contributing 
to $\Gamma_{ge}(\lambda)$, i.e., those corresponding to the transition $|e\rangle\rightarrow |g\rangle$, 
as relevant ones. The 
transition rate $\Gamma_{ge}(\lambda)$ is given by\cite{PhysRevB.90.165411} (see appendix Eq. \eqref{Gdiag2})
\begin{align}
  \Gamma_{ge}(\lambda)&= \cos^4\!\frac{\theta}{2} \gamma(\Omega\!+\!\omega)e^{i\lambda (\Omega+\omega)}
 \!\!+\!\sin^4\!\frac{\theta}{2} \gamma(\Omega\!-\!\omega)e^{i\lambda(\Omega-\omega)}.
\end{align}
The first term corresponds to processes in which a quantum of energy $\Omega + \omega$
is emitted to the bath. The second term describes events in which energy $\omega - \Omega$ is 
absorbed by the system from the bath.

At elevated temperatures, $T\sim \omega$, and at resonance,
$\theta = \pi/2$, both processes have comparable rates. Thus, on average, energy of order 
$\Omega$ is dissipated. Indeed, the cumulant $c_1(\pi,0)$ is of order 
$\Omega$ in this regime. In contrast, far from the resonance the first process dominates 
(this means that at any driving time one extra emission of $\omega+\Omega$ quantum has to occur)
and we obtain $c_1(\pi,0) \approx \omega+\Omega$, i.e., a much bigger energy than in the resonant regime. 
This also explains the enhancement of the emitted energy in the detuned
scenario for short driving times compared to the resonant situation as observed in \fig{fig:timeanalysis} (a).
Note that
within this regime the enhancement is way larger than the natural increase 
of the energy difference $\Omega = \sqrt{\Delta^2 +\Omega_R^2}$ due to the detuning.
The related effect of the detuning on the Mollow-triplet was discussed in Refs.~\cite{PhysRev.188.1969,Ulhaq:13,Saiko2014}.

Next, we analyze the selection sensitive time independent contribution $c_2(\theta_i,\theta_f)$
to the noise.
The quantity $c_2(\theta_i,\theta_f)$ is depicted in \fig{fig:constant_compare} for detuned driving 
(panel (c)) and at resonance (panel (d)). We again observe the qualitative similarity with the 
full numerical result presented in \fig{fig:variance_FI}. As in \fig{fig:variance_FI} the 
noise is enhanced and quite 
insensitive to the state selection in the regime of resonant driving. 
The structure for finite detuning (cf. \fig{fig:variance_FI} (a) and \fig{fig:constant_compare} (c)) appears to be more versatile 
and more sensitive to the state selection. 
Interestingly the noise turns out to be minimal in the vicinity of the $|e\rangle \rightarrow |g\rangle$ transition.
Indeed we find

\begin{align}
 c_2(\pi,0) &= \frac{\Gamma_{ge}''}{\Gamma_{ge}}- \left(\frac{\Gamma_{ge}'}{\Gamma_{ge}}\right)^2\nn\\
 &=\frac{\sin^4\theta  e^{\beta(\omega-\Omega)} \omega^2  \gamma(\omega-\Omega) \gamma(\omega+\Omega)}{4\left(\sin^4\frac{\theta}{2} \gamma(\omega-\Omega) +e^{\beta (\omega-\Omega)} 
 \cos^4\frac{\theta}{2} \gamma(\omega+\Omega) \right)^2},
\end{align}
which leads to a suppression of the conditional variance as the detuning 
increases due to the $\sin^4\theta$ dependency in the numerator. 
The physical explanation of this behavior of the selection dependent noise 
is similar to that of the time independent contribution to the conditional average. 
The state selection 
determines that the noise depends only on the corresponding rate of the master equation $\Gamma_{ge}$. When the system is driven resonantly, the $\Omega+\omega$ and $\Omega-\omega$
transitions appear to be equally like. Thus the noise is enhanced. In the detuned regime the $\Omega+\omega$ 
transition is favored and the noise is suppressed.

\section{Conclusion}
\label{sec:conclusion}

In this paper we study the effect on pre- and post-selection on the first two conditional cumulants of dissipated energy.
We report that not only the choice of initial and final states but additionally driving off 
resonance yields interesting and rich results.

The average heat current $\lim_{\tau\rightarrow \infty} \langle\epsilon_\tau\rangle_{i\rightarrow f}/\tau$ turns out to be independent of choice of the pre- and post selection
and  is only sensitive to the detuning of the drive. As one would expect, it becomes maximal when the 
driving is resonant.
For finite detuning and an energetically unfavorable choice of pre- and post-selected system states our analysis 
shows that the conditional average becomes negative at times of order of the relaxation times above a crossover temperature $T_0$. Further analysis shows 
that this temperature tends to diverge as a function of the driving time. Thus, for a long enough driving time
the system has to dissipate energy to the bath irrespective of how high its temperature is.

Furthermore we find that the state selection manifests itself mostly in a time independent contribution which turns out to
be sensitive to the detuning. 
In the vicinity of the $|e\rangle \rightarrow |g\rangle$ transition a detailed analysis shows that the increase of detuning favors
a distinct transition rate and therefore a distinct energy emission $\Omega + \omega$. 
This yields a suppression of the conditional noise.

As the effect is time-independent it may be most easily detectable after long driving times $\tau\gg \Gamma_\text{rel}^{-1}$ (as a small pre- and post-selection dependent correction to the selection independent contribution). 
At times of order of relaxation times, $\tau \sim \Gamma_\text{rel}^{-1}$, the selection dependent contribution 
may dominate. 

Furthermore our findings show that quantum corrections to the conditional average become more pronounced at lower temperatures and for pre- and post-selected system states with maximum coherence.

\section*{Acknowledgements}
We thank G. Sch\"on for valuable discussions. We acknowledge financial support
of the German Science Foundation (DFG Research Grant No. SH 81/2-1), the German-Israeli Foundation (GIF Research Grant No. 1183-229.14/2011) 
and JSPS KAKENHI (Grants No. 26400390 and No. JP26220711). 


\appendix
\begin{widetext}

\section{Characteristic function}
\label{sec:appendix_CF}
Within the Lindblad-Master equation approach, the characteristic function 
$\chi_\tau(\lambda,f|i) = \chi_\tau^p (\lambda,f|i) + \delta\chi_\tau(\lambda,f|i)$ separates\cite{PhysRevB.90.165411}
into a classical and a quantum part. The classical part, which is determined by the diagonal elements of the density matrix (populations) is given by
\begin{align}\label{eq:chi_p}
 \chi_\tau^p(\lambda,f|i) = \frac{e^{-\frac{\tau}{2}\left(\Gamma_{gg}(\lambda)+\Gamma_{ee}(\lambda) -2\Lambda(\lambda)\right)}}{2\Lambda(\lambda)} \left(\frac{A(\lambda,\theta_f,\theta_i)+B(\lambda,\theta_f,\theta_i)}{4} - \frac{A(\lambda,\theta_f,\theta_i)-B(\lambda,\theta_f,\theta_i)}{4} e^{-2\Lambda(\lambda)\tau} \right),
\end{align}
where 
\begin{align}
 A(\lambda,\theta_f,\theta_i)&= \left( \cos\theta_i +\cos \theta_f\right) (\Gamma_{ge}(0) -\Gamma_{eg}(0)) - (\cos \theta_f -1)(\cos\theta_i+1) \Gamma_{eg}(\lambda)
 -(\cos \theta_f+1) (\cos \theta_i-1) \Gamma_{ge}(\lambda)\label{eq:A}\\
 B(\lambda,\theta_f,\theta_i) &= 2\Lambda(\lambda ) (1+ \cos \theta_f \cos \theta_i)\label{eq:B}
\end{align}
and $\Gamma_{ij}(\lambda)$ are the transition rates. We also introduced 
$ \Lambda(\lambda) = \frac{1}{2}\sqrt{ 4\Gamma_{eg}(\lambda)  \Gamma_{ge}(\lambda) + (\Gamma_{ee}(\lambda)-\Gamma_{gg}(\lambda))^2 }$.
The quantum part  depends solely on the off-diagonal elements (coherences) and is given by
\begin{align}\label{eq:quantum_CF}
 \delta \chi_\tau(\lambda,f|i) = \frac{1}{2} \cos (\Omega \tau) \, \sin\theta_i \sin\theta_f e^{- \Gamma_\varphi(\lambda)\tau},
\end{align}
where $\Gamma_\varphi(\lambda)$ is the counting field dependent dephasing rate. 

The transition probability of finding the system after driving time $\tau$ in the desired final state $|f\rangle$ 
given the pre-selected initial state $|i\rangle$ is given by
\begin{align}
 \mathcal{P}_\tau(f|i) = \mathcal{P}^p_\tau(f|i) + \delta\mathcal{P}_\tau(f|i),
\end{align}
where
\begin{align}
\mathcal{P}^p_\tau(f|i)= \chi_\tau^p(0,f|i) = \frac{\Gamma_\text{rel} + \cos\theta_f (\Gamma_{ge}-\Gamma_{eg}) + e^{-\Gamma_\text{rel} t} 
\cos \theta_f (\cos\theta_i \Gamma_\text{rel} - \Gamma_{ge}+\Gamma_{eg})}{2\Gamma_\text{rel}},
\end{align}
and
\begin{align}
 \delta \mathcal{P}_\tau(f|i) = \delta \chi_\tau (0,f|i) = \frac{1}{2} \cos (\Omega t) \, \sin\theta_i \sin\theta_f e^{- \Gamma_\varphi(0)t}.
\end{align}
For the sake of readability we abbreviated $\Gamma_{ij}(0)\equiv \Gamma_{ij}$. 

The rates have been calculated in Ref.~\onlinecite{PhysRevB.90.165411} and are given by
\begin{align}
 \Gamma_{gg}(\lambda)&= \Gamma_{eg}(\lambda=0) -  \gamma^-(\omega,\lambda)\label{Gdiag1},\\
 \Gamma_{ge}(\lambda)&= \cos^4\frac{\theta}{2} \gamma(\Omega+\omega)e^{i\lambda (\Omega+\omega)}
 +\sin^4\frac{\theta}{2} \gamma(\Omega-\omega)e^{i\lambda(\Omega-\omega)}\label{Gdiag2} ,\\
 \Gamma_{eg}(\lambda)&=\cos^4\frac{\theta}{2}\gamma(-\Omega-\omega) e^{i\lambda(-\Omega-\omega)}
 +\sin^4\frac{\theta}{2} \gamma(-\Omega+\omega) e^{i\lambda(-\Omega+\omega)}\label{Gdiag3},\\
 \Gamma_{ee}(\lambda)&=\Gamma_{ge}(\lambda=0)-  \gamma^-(\omega,\lambda) \label{Gdiag4}\\
 \Gamma_\varphi(\lambda)&=   \gamma^+(\omega,\lambda) +\frac{1}{2}\left(\Gamma_{eg}(0)+\Gamma_{ge}(0)\right) \label{Gphi}.
\end{align}
Here 
\begin{align}\label{gammapm}
 \gamma^\pm(\omega,\lambda)&=\frac{\sin^2\theta}{4} \left(\gamma(\omega)\!\left(e^{i\lambda\omega} \pm 1\right) + \gamma(-\omega)\left( e^{-i\lambda\omega}\pm1\right)\right)
\end{align}
and $\gamma(\omega)= \int ds\, e^{i\omega s} \langle B(s)B(0)\rangle$ is the Fourier transform of the bath correlation functions.
\section{Asymptotic behavior of the conditional average}
\label{sec:appendix_asymptotics}
 For long enough driving times 
$\tau\gg \Gamma_\text{rel}^{-1},\Gamma_\varphi^{-1}$, the conditional average of dissipated energy grows
linear in the driving time. At such time scales the coherences already have died out. Hence, we can restrict
the analysis to the dynamics of the populations. 
We determine the average heat current as
\begin{align}
 \lim_{\tau\rightarrow \infty} \frac{\langle\epsilon_\tau\rangle_{i\rightarrow f}}{\tau} &= \lim_{\tau\rightarrow\infty} \frac{1}{\tau} \partial_{i\lambda} \ln \chi_\tau^p(\lambda,f|i)|_{\lambda=0}\nn\\
 &= \lim_{\tau\rightarrow \infty} \frac{1}{\tau}\partial_{i\lambda}\left.\left( \frac{\tau}{2}(-\Gamma_{gg}(\lambda)-\Gamma_{ee}(\lambda)+ 2\Lambda(\lambda) )
 -\ln 2\Lambda(\lambda) +\ln\frac{A+B}{4} + \ln\left( 1-\frac{A-B}{A+B}e^{-2\Lambda(\lambda)\tau} \right)\right)\right|_{\lambda=0}\nn\\
 &=\partial_{i\lambda} \frac{1}{2}\left(-\Gamma_{gg}(\lambda)-\Gamma_{ee}(\lambda)+ 2\Lambda(\lambda) \right)\nn\\
 &= \frac{\Gamma_{ge}' \Gamma_{eg} + \Gamma_{ge}\Gamma_{eg}'}{\Gamma_\text{rel}} - \frac{\Gamma_{gg}' +\Gamma_{ee}'}{2}\ ,
\end{align}
where $\Gamma_{ij}'\equiv \partial_{i\lambda} \Gamma|_{\lambda=0}$. In the regime $\omega > \Omega$
 we find the average heat current to be equal to
\begin{align}
 \lim_{\tau\rightarrow \infty} \frac{\langle\epsilon_\tau\rangle_{i\rightarrow f}}{\tau} = \frac{\sin^2\theta\,\omega}{4\Gamma_\text{rel}}
 \left(\Gamma_\text{rel} \gamma(\omega) \left(1-e^{-\beta\omega}\right)+ \frac{\sin^2\theta}{2} \gamma(\omega + \Omega) \gamma(\omega-\Omega) \left( 1-e^{-2\beta\omega}\right)\right),
\end{align}
where the angle $\theta$ is determined via $\tan \theta = \Omega_R/\Delta$. 
With this we immediately see that the largest heat current is achieved for $\Delta =0$, where the $\sin(\theta)=1$.


\end{widetext}

\bibliography{cond_av}{}

\end{document}